\documentclass[showpacs,showkeys,notitlepage,12pt]{revtex4}%
\usepackage{amssymb}
\usepackage{amsfonts}
\usepackage{amsmath}
\usepackage{latexsym}
\usepackage{graphicx}%
\begin{document}
\title{Stability of a class of neutral vacuum bubbles\\ }
\author{J.R. Morris}
\affiliation{Physics Dept., Indiana University Northwest, 3400 Broadway, Gary, Indiana
46408 USA}

\begin{abstract}
A model that gives rise to vacuum bubbles is considered where the domain wall
field interacts with another real scalar field, resulting in the formation of
domain ribbons within the host domain wall. Ribbon-antiribbon annihilations
produce elementary bosons whose mass inside the wall is different from the
mass in vacuum. Two cases are considered, where the bosons get trapped either
within the bubble wall or the bosons get trapped within the vacuum enclosed by
the bubble. The bosonic (meta)stabilization effect on the bubble is examined
in each case. It is found that when the bosons become trapped within the
bubble wall, the stabilization mechanism lasts for only a limited amount of
time, and then the bubble undergoes unchecked collapse. However, when the
bosons become trapped within the bubble's interior volume, the bubble can be
long-lived, provided that it has a sufficiently thin wall.

\end{abstract}

\maketitle

\section{Introduction}

\bigskip

\ \ The stabilizing effects of fermions on vacuum bubbles and other
nontopological solitons have been studied in a number of interesting field
theoretical scenarios\cite{Sahu}-\cite{JM98}. In addition, models have been
examined where such objects may be stabilized by bosons which can get trapped
within the defect, due to a mass contrast between interior and exterior
regions\cite{Coleman}-\cite{JM04}. Presently, we consider models for possible
bosonic stabilizing effects, where elementary bosons are produced by
annihilations of defects (\textquotedblleft domain ribbons\textquotedblright)
residing within the host defect (domain wall). This is similar in nature to a
recent study investigating the possibility of a boson gas (meta)stabilization
of cosmic string loops\cite{JM13}.

\bigskip

\ \ The domain wall is described by a real scalar field $\chi$ which
interpolates between true vacuum states $\chi=\pm\eta$, but the field $\chi$
also interacts with a second scalar field $\phi$, as described in
Ref.\cite{Morris95}. By the Witten mechanism\cite{Witten}, there is a certain
range of model parameters allowing the scalar $\phi$ to settle into nonzero
vacuum states within the core of the wall, taking on values where $\phi
=\pm\phi_{0}$, thereby breaking a discrete $Z_{2}$ symmetry associated with
the field $\phi$ within the wall. However, the $+\phi_{0}$ and $-\phi_{0}$
vacuum domains will be uncorrelated beyond some coherence length $\xi$, and a
$+\phi_{0}$ domain and an adjacent $-\phi_{0}$ domain will be separated by
region where $\phi=0$, which locates the center of a domain \textquotedblleft
ribbon\textquotedblright\ or \textquotedblleft antiribbon\textquotedblright%
\ described by $\phi_{R}$ and $\phi_{\bar{R}}$. So, a ribbon or antiribbon is
a topological domain wall section trapped within the host domain wall formed
by the field $\chi$. A single ribbon solution interpolates between the vacuum
states $\pm\phi_{0}$, and two adjacent ribbons are separated by an antiribbon,
and vice versa.

\bigskip

\ \ The ribbons and antiribbons are dynamical objects\ which can interact with
each other and themselves, undergoing annihilations and formations of ribbon
loops through fission and fusion processes. A ribbon loop is surrounded by a
$\pm\phi_{0}$ domain and encloses a $\mp\phi_{0}$ domain. A section of ribbon
(R) and a section of antiribbon (\={R}) can undergo annihilation, resulting in
the formation of elementary $\varphi$ bosons, which are the perturbative
particle excitations of the $\phi$ field. The $\varphi$ bosons have a mass
$m_{\text{in}}$ inside the domain wall and a mass $m$ outside the wall in a
true vacuum. If there is a high mass contrast between $m_{\text{in}}$ and $m$,
then $\varphi$ bosons will tend to either become trapped within the wall
($m_{\text{in}}\ll m$) or be expelled out of the wall ($m_{\text{in}}\gg m$).
A formation of $\chi$ vacuum bubbles can result, due to wall-wall interactions
or self-intersecting trajectories. Also, if the $Z_{2}$ symmetry with
degenerate vacuum states $\chi=\pm\eta$ is biased with different probabilities
of forming different domains becoming unequal\cite{bias}, or if the $Z_{2}$
symmetry is approximate with a negligible difference in vacuum
energies\cite{approximate}, a network of bounded domain wall surfaces may
result, leading to vacuum bubbles. The $\varphi$ particles which either get
trapped within the bubble wall or the interior volume of the bubble can then
have a (meta)stabilizing effect. Although the issue of long term bubble
stability is difficult to address with confidence, the shorter term bosonic
stabilizing effect can be analyzed for both cases $m_{\text{in}}\ll m$ and
$m_{\text{in}}\gg m$.

\bigskip

\ \ The basic domain ribbon model is described in Sec. II. Bosonic
stabilization is then analyzed in Sec.III for the case $m_{\text{in}}\ll m$
and in Sec. IV for the case $m_{\text{in}}\gg m$. Equilibrium bubble radii and
bubble masses are obtained for each case, and the possibility of long term
stability is considered. It is concluded that for the case where $\varphi$
bosons become trapped within the bubble wall ($m_{\text{in}}\ll m$), the
bosonic stabilization mechanism is initially effective, with a slow rate of
leakage of $\varphi$ gas due to a small number of bosons with energies $E\geq
m$ escaping the wall. However, the gas temperature subsequently rises, and the
stability mechanism eventually becomes ineffective, with much of the $\varphi$
boson gas escaping the bubble wall. On the other hand, for the case where the
$\varphi$ gas is trapped within the bubble's interior volume ($m_{\text{in}%
}\gg m$), there is a possibility for long term stability, provided that the
bubble wall is thin enough. A brief summary is presented in Sec. V.

\section{Domain ribbon model}

\ \ The Lagrangian for the system of two interacting scalar fields $\chi$ and
$\phi$ is\cite{Morris95}%
\begin{equation}
\mathcal{L}=\frac{1}{2}\partial^{\mu}\chi\partial_{\mu}\chi+\frac{1}%
{2}\partial^{\mu}\phi\partial_{\mu}\phi-V\left(  \chi,\phi\right)  \label{1}%
\end{equation}

where the potential is%
\begin{equation}
V\left(  \chi,\phi\right)  =\frac{1}{4}\lambda\left(  \chi^{2}-\eta
^{2}\right)  ^{2}+\frac{1}{2}f\left(  \chi^{2}-\eta^{2}\right)  \phi^{2}%
+\frac{1}{2}m^{2}\phi^{2}+\frac{1}{4}g\phi^{4} \label{2}%
\end{equation}

The stable vacuum states of this system are located by $\chi=\pm\eta$,
$\phi=0$ The parameters $\lambda$, $f$, $\eta$, $m$, and $g$ are taken to be
real-valued and positive. The field equations obtained from $\mathcal{L}$ are%
\begin{equation}
\nabla_{\mu}\partial^{\mu}\chi+\left[  \lambda\left(  \chi^{2}-\eta
^{2}\right)  +f\phi^{2}\right]  \chi=0 \label{3}%
\end{equation}

\begin{equation}
\nabla_{\mu}\partial^{\mu}\phi+\left[  f\left(  \chi^{2}-\eta^{2}\right)
+m^{2}+g\phi^{2}\right]  \phi=0 \label{4}%
\end{equation}

\ \ We notice that for $\phi=0$ the field equation for $\chi$ admits a domain
wall solution describing a planar domain wall centered on the $y-z$ plane:%
\begin{equation}
\chi(x)=\eta\tanh\left(  \frac{x}{w}\right)  ,\,\,\,\,\,\,\,\,w=\frac{1}{\eta
}\sqrt{\frac{2}{\lambda}}\,\, \label{5}%
\end{equation}

where $w$ is the thickness of the wall. Now, if the field $\phi$ does not
vanish identically, the Witten mechanism\cite{Witten} allows a $\phi$
condensate to form within the wall taking values $\phi=\pm\phi_{0}$, where%

\begin{equation}
\phi_{0}=\left[  \left(  f\eta^{2}-m^{2}\right)  /g\right]  ^{\frac{1}{2}}>0
\label{6}%
\end{equation}

\ \ For mathematical simplicity and approximation purposes we take the wall to
be a slab of thickness $w$, with%

\begin{equation}
\left\vert \chi\right\vert \approx\left\{
\begin{array}
[c]{cc}%
0, & \left\vert x\right\vert \leq\frac{1}{2}w\\
\eta, & \left\vert x\right\vert >\frac{1}{2}w
\end{array}
\right\}  ,\,\,\,\,\,\,\,\,\,\,\phi\approx\left\{
\begin{array}
[c]{cc}%
\phi(y,z,t), & \left\vert x\right\vert \leq\frac{1}{2}w\\
0, & \left\vert x\right\vert >\frac{1}{2}w
\end{array}
\right\}  . \label{7}%
\end{equation}

\noindent\ Then (\ref{4}), with the help of (\ref{7}), allows us to express
the equation of motion for $\phi$ inside the domain wall by, approximately,%

\begin{equation}
\partial_{0}^{2}\phi-\left(  \partial_{y}^{2}\phi+\partial_{z}^{2}\phi\right)
+g\phi\left(  \phi^{2}-\phi_{0}^{2}\right)  =0 \label{8}%
\end{equation}

Equation (\ref{8})\noindent\ admits the static solution%

\begin{equation}
\phi_{R}(z)=\phi_{0}\tanh\left(  \frac{z-z_{0}}{w_{R}}\right)
,\,\,\,\,\,\,\,\,\,\,w_{R}=\frac{1}{\phi_{0}}\sqrt{\frac{2}{g}} \label{9}%
\end{equation}

\noindent describing a \textit{domain ribbon} of width $w_{R}$ embedded within
the wall, lying along the $y$ direction with $\phi_{R}\rightarrow\pm\phi_{0}$
as $z\rightarrow\pm\infty$. The ribbon (R) thus separates $\pm\phi_{0}$
domains, and the antiribbon (\={R}) solution is given by $\phi_{\bar{R}%
}(z)=-\phi_{R}(z)$. We picture the ribbon as a section of domain wall, lying
on the $y$ axis, with thickness $w_{R}$ in the $z$ direction and thickness $w$
in the $x$ direction.

\bigskip

\ \ In the domain wall background described by (\ref{7}), we see from
(\ref{2}) and (\ref{6}) that $\phi=0$ is not the lowest energy state inside
the $\chi$ wall, but rather $\phi=\pm\phi_{0}$ is. Therefore, a $\phi$ field
forms a condensate within the wall, tending to settle into either $\phi
=+\phi_{0}$ or $\phi=-\phi_{0}$ domains, but these domains will be
uncorrelated beyond \ some coherence length $\xi\gtrsim w_{R}\sim1/(\sqrt
{g}\phi_{0})$. Two different adjacent domains must be separated by a ribbon or
antiribbon located where $\phi=0$. More general solutions of (\ref{8}) would
include infinite ribbons, which need not be straight or static, and closed
(anti)ribbon loops\cite{Morris95}. These ribbon configurations therefore
resemble cosmic strings. A closed loop is surrounded by a $\pm\phi_{0}$
domain, and encloses a $\mp\phi_{0}$ domain. Smaller loops can fuse together
to form larger loops, and larger loops can fission by self-intersection to
form smaller loops. These processes involve R-\={R} annihilations wherever the
loops intersect, resulting in the release of $\varphi$ boson radiation, where
the $\varphi$ boson is the elementary particle excitation of the $\phi$ field
inside the $\chi$ wall, i.e., $\phi=\phi_{0}+\varphi$, for example. For our
model, we take the $\varphi$ particles to be much less massive than loops,
which have masses $M_{\text{loop}}=\mu_{R}L$, where $\mu_{R}$ is the linear
energy density of a ribbon (tension), and $L$ is the loop length. Therefore,
we assume R-\={R} annihilations resulting in $\varphi$ production to be much
more likely than loop creation due to $\varphi$ scattering.

\bigskip

\ \ The $\varphi$ particle mass inside the $\chi$ wall is denoted by
$m_{\text{in}}$, while the $\phi$ particle mass outside the wall is denoted by
$m$. There are two possibilities we wish to consider.

\bigskip

\ \ (1) $m_{\text{in}}\ll m$, in which case $\varphi$ particles get trapped
within the $\chi$ wall, forming a boson gas at temperature $T=1/\beta$ with a
thermal distribution of particle energies. In this case only particles with
energies $E\geq m$ can escape from the wall to the vacuum outside. If $T\ll
m$, i.e., $\beta m\gg1$, then the number of particles $N(E)=(e^{\beta
E}-1)^{-1}$ with enough energy to escape is small, and the $\varphi$ leakage
rate is low, and the gas pressure tends to counteract the $\chi$ wall tension,
allowing a spherical bubble to find equilibrium at a finite radius $R$.
Bubbles of this scenario are dubbed \textquotedblleft type 1\textquotedblright\ bubbles.

\bigskip

\ \ (2) $m_{\text{in}}\gg m$, and in this case $\varphi$ particles are
expelled out of the $\chi$ wall and into surrounding vacuum. However, $\chi$
walls are themselves dynamic and can form closed bubbles through mechanisms
mentioned earlier. If $\chi$ bubbles form before R-\={R} annihilations
complete, a $\varphi$ gas will become trapped in the interior of the bubble,
and exert an outward pressure tending to counteract the effects of bubble
tension, allowing an equilibrium state for the bubble. Bubbles of this
scenario are dubbed \textquotedblleft type 2\textquotedblright\ bubbles.

\section{Type 1 bubble stability}

\subsection{Bosonic Stabilization}

\ \ For the type 1 bubble $m_{\text{in}}\ll m$ and a gas of $\varphi$ bosons
is trapped inside the $\chi$ bubble wall, which we assume to take a spherical
shape. Dynamical bubbles can emit radiation in the form of $\chi$ particles of
mass $m_{\chi}=\sqrt{2\lambda}\eta$. Now consider a bubble which is shrinking
under the influence of its wall tension. As it shrinks, the amount of bubble
wall volume available to the $\varphi$ gas decreases, and we expect the
temperature of the gas to rise. We assume that the $\varphi$ gas is
relativistic, with effectively massless particles, so that a nonnegligible
energy density $\rho_{G}$, number density $n$, and pressure $p$ can exist to
play a role in a stabilization of the bubble. (We ignore possible effects of
$\chi$ particles emitted into the bubble's interior vacuum, as they can pass
right through the $\chi$ wall - the reflection coefficient is
zero\cite{section} - and pass into the exterior vacuum. Similarly, we ignore
the possibility of $\varphi$ bosons inside of a closed ribbon loop exerting an
outward pressure on the loop, since reflectionless scattering of $\varphi$
particles from a $\phi$ wall is expected.) The radial force acting on the
bubble wall is $F_{R}=-\partial\mathcal{E}/\partial R$, where $\mathcal{E}$ is
the configuration energy of the nontopological soliton composed of vacuum
bubble and boson gas. If $F_{R}<0$ the bubble shrinks, and for $F_{R}>0$ the
bubble tends to expand at the expense of $\varphi$ particle energy, and at
equilibrium the configuration energy $\mathcal{E}$ is minimized with $F_{R}=0$.

\bigskip

\ \ The configuration energy $\mathcal{E}$ of the bubble is the sum of two
terms, the energy $\mathcal{E}_{W}$ of the bubble wall due to wall surface
energy density $\Sigma$ (tension), and the energy $\mathcal{E}_{G}$ of the
relativistic $\varphi$ boson gas trapped within the wall, which has a volume
of $w(4\pi R^{2})$, where $R$ is the bubble radius. Then%
\begin{equation}
\mathcal{E}=\mathcal{E}_{W}+\mathcal{E}_{G}=\Sigma(4\pi R^{2})+\rho_{G}(4\pi
R^{2}w)=4\pi R^{2}\left(  \Sigma+\frac{\pi^{2}}{30}T^{4}w\right)  \label{10}%
\end{equation}

where the $\varphi$ gas energy density is $\rho_{G}=\frac{\pi^{2}}{30}T^{4}$.
The equilibrium radius $R$ is determined by minimizing $\mathcal{E}$ with
respect to $R$. We must be careful, however, in this minimization. We want the
minimal value of $\mathcal{E}$ for a given $\varphi$ particle number $N$ and a
given entropy $S$. Therefore we consider a \textit{virtual} variation of
$\mathcal{E}$ holding $N$ and $S$ fixed. (One condition implies the other,
since both are proportional to $T^{3}(4\pi R^{2}w)$.) The number density of
$\varphi$ particles at temperature $T$ is\cite{KTbook} $n=\frac{\zeta(3)}%
{\pi^{2}}T^{3}$ and the entropy density is\cite{KTbook} $s=\frac{2\pi^{2}}%
{45}T^{3}$. We then have%
\begin{equation}
N=n(4\pi R^{2}w)=\frac{4w\zeta(3)}{\pi}T^{3}R^{2},\ \ \ \ \ S=s(4\pi
R^{2}w)=\frac{8\pi^{3}w}{45}T^{3}R^{2} \label{11}%
\end{equation}

so that holding $N$ and S constant during the virtual variation of
$\mathcal{E}$ implies the constraint%
\begin{equation}
T^{3}R^{2}=\frac{N\pi}{4w\zeta(3)}=\frac{45S}{8\pi^{3}w}\equiv C^{3}%
;\ \ \ \ T=\frac{C}{R^{2/3}};\ \ \ \ \ \ \ \ C=\left[  \frac{N\pi}{4w\zeta
(3)}\right]  ^{1/3}=\left[  \frac{45S}{8\pi^{3}w}\right]  ^{1/3} \label{12}%
\end{equation}

Using the relation $T=CR^{-2/3}$ in (\ref{10}) then gives%
\begin{equation}
\frac{\mathcal{E}}{4\pi}=\Sigma R^{2}+\frac{\pi^{2}wC^{4}}{30}R^{-2/3}
\label{13}%
\end{equation}

Minimizing this expression by requiring $\partial\mathcal{E}/\partial R=0$
results in an equilibrium radius given by%
\begin{equation}
R^{8/3}=\frac{\pi^{2}wC^{4}}{90\Sigma};\ \ \ \ \ R=\left[  \frac{\pi^{2}%
wC^{4}}{90\Sigma}\right]  ^{3/8} \label{14}%
\end{equation}

\bigskip

\ \ Equations (\ref{13}) and (\ref{14}) then give the bubble mass at
equilibrium,%
\begin{equation}
\frac{\mathcal{E}}{4\pi}=\frac{\mathcal{E}_{W}}{4\pi}+\frac{\mathcal{E}_{G}%
}{4\pi};\ \ \mathcal{E}_{W}=\Sigma R^{2},\ \ \ \ \mathcal{E}_{G}=\frac{\pi
^{2}wC^{4}}{30}R^{-2/3} \label{15}%
\end{equation}

Upon comparing the two energy terms we find, with the help of (\ref{14}),%
\begin{equation}
\frac{\mathcal{E}_{G}}{\mathcal{E}_{W}}=3;\ \ \ \mathcal{E}=\mathcal{E}%
_{W}+\mathcal{E}_{G}=4\mathcal{E}_{W}=\frac{4}{3}\mathcal{E}_{G}=16\pi\Sigma
R^{2} \label{16}%
\end{equation}

so that a type 1 bubble of radius $R$ at equilibrium has a mass $\mathcal{E}%
=16\pi\Sigma R^{2}$. From (\ref{16}) we also have $\mathcal{E}_{G}%
=3\mathcal{E}_{W}$, which leads to [see, e.g., (\ref{10})] an equilibrium
temperature given by%
\begin{equation}
\frac{\pi^{2}}{30}T^{4}w=3\Sigma\ \ \implies\ \ T=\left(  \frac{90\Sigma}%
{\pi^{2}w}\right)  ^{1/4} \label{16a}%
\end{equation}

Therefore, all type 1 bubbles equilibrate at this temperature, regardless of size.

\subsection{Bubble Decay}

\ \ Even with a bosonic stabilization mechanism, we don't expect the bubble to
remain in equilibrium forever. This is because high energy $\varphi$ particles
with energies $E\geq m$ are energetically allowed to escape the bubble wall
and eventually end up in the exterior vacuum. Although $T\ll m$ for a bubble
near equilibrium, we expect a rate of leakage which is slow at first, but then
increases due to an increase in temperature and average $\varphi$ particle
energy. A dynamical bubble can also shrink through emission of $\chi$
particles. Let us focus on a simple scenario of type 1 bubble decay due to
$\varphi$ particle loss. We make the simple assumption that each $\varphi$
particle that escapes the bubble becomes a $\phi$ particle of mass $m$ outside
the bubble. Suppose the bubble is initially in a state near equilibrium with
mass $\mathcal{E}_{0}$ and $\varphi$ particle number $N_{0}$ at some initial
time $t_{0}$. After a time $t-t_{0}$ the bubble has a mass $\mathcal{E}$ and
$\varphi$ particle number $N$, and has emitted $N_{\phi}=N_{0}-N$ particles
into the exterior vacuum. Then we can write $\mathcal{E}_{0}=\mathcal{E}%
+N_{\phi}m+\mathcal{E}_{X}$, where $\mathcal{E}_{X}$ includes $\phi$ boson
kinetic energy and $\chi$ radiation energy. The energy $\mathcal{E}_{X}$ will
be a positive, monotonically increasing function of time $t$. Now let us
rewrite this in the form%
\begin{equation}
Nm-\mathcal{E}=K+\mathcal{E}_{X}\equiv Q(t) \label{17}%
\end{equation}

where $K=N_{0}m-\mathcal{E}_{0}$ and $Q(t)$ is a positive, monotonically
increasing function. We see that $K>0$ since the bubble is initially close to
equilibrium and $N_{0}m>\mathcal{E}_{0}$. To be more explicit, let us write%
\begin{equation}
\frac{N}{4\pi R^{2}}=\frac{\zeta(3)wT^{3}}{\pi^{2}}\equiv DT^{3}%
,\ \ \ \ \ \frac{\mathcal{E}_{G}}{4\pi R^{2}}=\frac{\pi^{2}w}{30}T^{4}\equiv
BT^{4} \label{18}%
\end{equation}

where%
\begin{equation}
D=\frac{\zeta(3)w}{\pi^{2}},\ \ B=\frac{\pi^{2}w}{30},\ \ \ \frac{B}%
{D}=2.7,\ \ \ \ \frac{D}{B}=.37 \label{19}%
\end{equation}

As long as the bubble remains sufficiently close to equilibrium, with $T\ll
m$, we can write (\ref{17}) in the form%
\begin{equation}
mDT^{3}-BT^{4}=\Sigma+\frac{Q}{4\pi R^{2}} \label{20}%
\end{equation}

The first term on the left hand side dominates the second (and $K>0$ for this
same reason \cite{note}), so that we have, approximately,%
\begin{equation}
T\approx\left[  \frac{1}{mD}\left(  \Sigma+\frac{Q}{4\pi R^{2}}\right)
\right]  ^{1/3} \label{21}%
\end{equation}

for times where the bubble does not stray too far from equilibrium. As the
bubble shrinks, the $\varphi$ temperature increases, and the leakage rate increases.

\bigskip

\ \ Eventually, the bubble will have evolved too far away from an equilibrium
state for the approximations above to remain valid. This is seen by noticing
that at high enough temperatures the left hand side of (\ref{20}) begins
decreasing, while the right hand side is increasing. This happens for
$T\gtrsim T_{m}$, where%
\begin{equation}
T_{m}=\frac{3D}{4B}m\approx\frac{1}{4}m \label{22}%
\end{equation}

locates the local maximum of the left hand side of (\ref{20}). Upon
approaching this temperature, we must assume that the bubble rapidly loses its
$\varphi$ gas, and undergoes an unchecked collapse. In fact, at the
temperature $T_{m}$ we have $\beta E\geq\beta m$ for particles energetic
enough to escape the bubble wall, with $\beta m=m/T_{m}\approx4$. The energy
density of the portion of the gas with particles with energy $E\geq m$ is then
given, approximately, by\cite{Reif}%
\begin{equation}
u(T_{m})=\frac{T_{m}^{4}}{2\pi^{2}}\int_{\beta m=4}^{\infty}\frac{\eta^{3}%
}{e^{\eta}-1}d\eta=\frac{T_{m}^{4}}{2\pi^{2}}I(4) \label{23}%
\end{equation}

where $\eta=\beta E$ and $I(4)=2.6$. The energy density of the entire boson
gas is $\rho_{G}(T_{m})=\frac{T_{m}^{4}}{2\pi^{2}}I(0)$ (taking the $\varphi$
particles to be effectively massless, with $m_{\text{in}}\sim0$), so that a
comparison gives%
\begin{equation}
\frac{u(T_{m})}{\rho_{G}(T_{m})}=\frac{I(4)}{I(0)}=\frac{2.6}{\pi^{4}/15}=.4
\label{24}%
\end{equation}

So at $T\sim T_{m}$ roughly 40\% of the bosonic gas has been lost, and the
stabilization mechanism rapidly comes to an end. On the other hand, for $T\ll
m$, i.e., $\beta m\gg1$, we have%
\begin{equation}
I(\beta m)=\int_{\beta m}^{\infty}\frac{\eta^{3}}{e^{\eta}-1}d\eta\approx
\int_{\beta m}^{\infty}\eta^{3}e^{-\eta}d\eta\approx(\beta m)^{3}e^{-\beta
m}\ll1 \label{25}%
\end{equation}

so that $u(T)/\rho_{G}(T)\ll1$ when the bubble is very close to equilibrium.
Therefore the $\varphi$ leakage rate is very low intially, but rapidly
increases, so that the bosonic stabilization mechanism operates effectively
for only a limited time span. After that, the type 1 bubble collapses.

\section{Type 2 bubble stability}

\subsection{Bosonic Stabilization}

\ \ For the type 2 bubble $m_{\text{in}}\gg m$ and $\varphi$ particles that
are produced within the bubble wall are accelerated out of the wall into the
vacuum as much lighter $\phi$ bosons. However, if the dynamical $\chi$ walls
form bubbles while R-\={R} annihilations are in progress, some of the $\phi$
particles will become trapped within the volume of the bubble's interior, and
will produce an outward pressure that has a stabilizing effect on the bubble.
We assume the gas of light $\phi$ particles to be relativistic, with
nonnegligible energy density $\rho_{G}$, number density $n$, and pressure $p$.
Again, the radial force acting on the bubble wall is $F_{R}=-\partial
\mathcal{E}/\partial R$, where $\mathcal{E}$ is the configuration energy of
the nontopological soliton composed of the $\chi$ bubble wall and the enclosed
$\phi$ gas. An equilibrium state exists when $\mathcal{E}$ is minimized for
some radius $R$.

\bigskip

\ \ There are again two contributions to the energy $\mathcal{E}$, one from
the bubble wall and one from the enclosed $\phi$ gas:%
\begin{equation}
\mathcal{E}=\mathcal{E}_{W}+\mathcal{E}_{G}=\Sigma(4\pi R^{2})+\rho_{G}\left(
\frac{4}{3}\pi R^{3}\right)  =4\pi\left(  \Sigma R^{2}+\frac{\pi^{2}}{90}%
T^{4}R^{3}\right)  \label{26}%
\end{equation}

where again $\rho_{G}=\frac{\pi^{2}}{30}T^{4}$. Again, we perform a virtual
variation of $\mathcal{E}$ with respect to $R$ while holding the $\phi$
particle number $N$ and $\phi$ entropy $S$ fixed. Again using $n=\frac
{\zeta(3)}{\pi^{2}}T^{3}$ for the $\phi$ particle number density and
$s=\frac{2\pi^{2}}{45}T^{3}$ for the $\phi$ entropy, we have%
\begin{equation}
N=n\left(  \frac{4}{3}\pi R^{3}\right)  =\frac{4\zeta(3)}{3\pi}T^{3}%
R^{3},\ \ \ \ \ S=s\left(  \frac{4}{3}\pi R^{3}\right)  =\frac{8\pi^{3}}%
{135}T^{3}R^{3} \label{27}%
\end{equation}

Requiring $N$ and $S$ to remain fixed during the virtual variation results in
the constraint%
\begin{equation}
T^{3}R^{3}=\frac{3\pi N}{4\zeta(3)}=\frac{135S}{8\pi^{3}}\equiv C_{0}%
^{3};\ \ \ \ T=\frac{C_{0}}{R};\ \ \ \ \ C_{0}=\left[  \frac{3\pi N}%
{4\zeta(3)}\right]  ^{1/3}=\left[  \frac{135S}{8\pi^{3}}\right]  ^{1/3}
\label{28}%
\end{equation}

Using the constraint $T=C_{0}/R$ in (\ref{26}) results in%
\begin{equation}
\frac{\mathcal{E}}{4\pi}=\Sigma R^{2}+\frac{\pi^{2}C_{0}^{4}}{90R} \label{29}%
\end{equation}

The equilibrium condition $\partial\mathcal{E}/\partial R=0$ yields an
equilibrium bubble radius%
\begin{equation}
R=\left(  \frac{\pi^{2}C_{0}^{4}}{180\Sigma}\right)  ^{1/3}\equiv\left(
\frac{C_{1}}{\Sigma}\right)  ^{1/3};\ \ \ C_{1}=\frac{\pi^{2}C_{0}^{4}}{180}
\label{30}%
\end{equation}

we can then write%
\begin{equation}
\frac{\mathcal{E}_{W}}{4\pi}=C_{1}^{2/3}\Sigma^{1/3},\ \ \ \frac
{\mathcal{E}_{G}}{4\pi}=2C_{1}^{2/3}\Sigma^{1/3}=2\frac{\mathcal{E}_{W}}{4\pi
},\ \ \ \ \frac{\mathcal{E}_{G}}{\mathcal{E}_{W}}=2 \label{31}%
\end{equation}

Therefore, the mass of the type 2 bubble, when in equilibrium, is given by%
\begin{equation}
\mathcal{E}=\mathcal{E}_{W}+\mathcal{E}_{G}=3\mathcal{E}_{W}=\frac{3}%
{2}\mathcal{E}_{G}=12\pi\Sigma R^{2} \label{32}%
\end{equation}

From $\mathcal{E}_{G}=2\mathcal{E}_{W}$ it follows that at equilibrium the
temperature of the $\phi$ gas is given by%
\begin{equation}
\left(  \frac{\pi^{2}}{30}T^{4}\right)  \frac{4}{3}\pi R^{3}=\Sigma\left(
8\pi R^{2}\right)  \ \implies\ T=\left(  \frac{180\Sigma}{\pi^{2}R}\right)
^{1/4} \label{33}%
\end{equation}

indicating that at equilibrium, larger bubbles have a lower temperature than
smaller ones.

\subsection{Bubble Decay}

\ \ We would like to know under what conditions a type 2 bubble can remain in
a long-lived near-equilibrium state. A long lived bubble must not leak $\phi$
particles to the outside vacuum at a significant rate, which requires that the
reflection coefficient $\mathcal{R}$ of the bubble wall be near unity. For
this condition to be met, the energies of most of the $\phi$ particles must
satisfy\cite{section} $E\ll E_{c}\equiv w^{-1}$, where we have introduced an
energy $E_{c}=w^{-1}$, with $w$ being the thickness of the bubble wall. For a
typical $\phi$ particle energy $E\sim T$, (and therefore $\beta E\sim1$) we
then require $\beta E\ll\beta E_{c}$, i.e., $\beta E_{c}\gg1$, to have
$\mathcal{R}\approx1$. Furthermore, under this condition the number of
particles with such low energy will be $N(E)=(e^{\beta E}-1)^{-1}\gg N(E_{c}%
)$, so that most $\phi$ particles will be reflected from the wall and remain
inside the bubble. Another way to see this is to look at the fraction of the
total energy density of the gas for particles having energies $E\geq E_{c}$.
The energy density associated with this set of particle energies is given
by\cite{Reif}%
\begin{equation}
u(T;\beta E_{c})=\frac{T^{4}}{2\pi^{2}}\int_{\beta E_{c}}^{\infty}\frac
{\eta^{3}d\eta}{e^{\eta}-1}=\frac{T^{4}}{2\pi^{2}}I(\beta E_{c}) \label{34}%
\end{equation}

where $\eta=\beta E$. Now, for $\beta E_{c}\gg1$ we have%
\begin{equation}
I(\beta E_{c})\approx\int_{\beta E_{c}}^{\infty}\eta^{3}e^{-\eta}d\eta
\approx(\beta E_{c})^{3}e^{-\beta E_{c}}\ll1,\ \ \ \ (\beta E_{c}\gg1)
\label{35}%
\end{equation}

The total energy density of the $\phi$ gas is $\rho_{G}=u(T;0)=\frac{T^{4}%
}{2\pi^{2}}I(0)$, where\cite{Reif} $I(0)=\pi^{4}/15$. We therefore have%
\begin{equation}
\frac{u(T;\beta E_{c})}{\rho_{G}}=\frac{I(\beta E_{c})}{I(0)}\approx\left(
\frac{15}{\pi^{4}}\right)  (\beta E_{c})^{3}e^{-\beta E_{c}}\ll
1,\ \ \ \ (\text{for\ }\beta E_{c}\gg1) \label{36}%
\end{equation}

Therefore, for the bosonic stabilization mechanism to be long-lived, with a
very low $\phi$ leakage rate, we require $\beta E_{c}\gg1$, i.e.,%
\begin{equation}
T\ll w^{-1} \label{37}%
\end{equation}

With a bubble remaining near equilibrium, this condition, along with
(\ref{33}) implies that the bubble radius $R$ satisfy%
\begin{equation}
R=\frac{180\Sigma}{\pi^{2}T^{4}}\gg\Sigma w^{4} \label{38}%
\end{equation}

Therefore, larger equilibrium bubbles, with thin walls, will decay at a slower
rate, and be longer lived than smaller ones.

\section{Summary}

\ \ A model of two interacting real scalar fields $\chi$ and $\phi$ has been
considered where the spontaneous breaking of a $Z_{2}$ symmetry associated
with the field $\chi$ gives rise to domain walls, which are dynamical, and can
interact to form closed bubbles. Within the core of a $\chi$ wall the $\phi$
field will tend to settle into ground states $\phi=\pm\phi_{0}$, breaking a
$Z_{2}$ symmetry associated with $\phi$. However, the $+\phi_{0}$ and
$-\phi_{0}$ domains will be uncorrelated beyond some coherence length $\xi$,
and adjacent domains wil be separated by a domain \textquotedblleft
ribbon\textquotedblright\ (R) or \textquotedblleft
antiribbon\textquotedblright\ (\={R}). These ribbon structures, confined to
the core of the $\chi$ wall, can behave as cosmic strings would, and exist
either as long wiggly objects or in the form of closed loops.

\bigskip

\ \ There will be R-\={R} annihilations taking place which result in the
production of $\varphi$ bosons, which are the elementary quanta of the
$\pm\phi_{0}$ vacuua. If these $\varphi$ bosons are produced within a bubble
wall, they can get trapped within the wall, if $m_{\text{in}}\ll m$, where
$m_{\text{in}}$ is the mass of the $\varphi$ particles within the wall, where
$\phi=\pm\phi_{0}+\varphi$, and $m$ is the $\phi$ particle mass in true vacuum
outside the wall, where $\phi=0$ locates the vacuum. These trapped $\varphi$
bosons form a boson gas within the wall with pressure $p$, which tends to
counteract the effects of the bubble wall tension $\Sigma$. On the other hand,
if $m_{\text{in}}\gg m$, the $\varphi$ particles will be expelled from the
wall, appearing as $\phi$ particles in vacuum. Some of these $\phi$ particles
will get trapped within the bubble's enclosed volume, and unless a substantial
fraction of these $\phi$ particles have a high enough energy to be transmitted
through the wall, they will form a boson gas which exerts an outward pressure
$p$, again, having a tendency to counteract the effects of the wall tension
$\Sigma$. Bubbles of the former case, where $m_{\text{in}}\ll m$, have been
dubbed \textquotedblleft type 1\textquotedblright\ bubbles, and bubbles of the
latter case, where $m_{\text{in}}\gg m$ are dubbed \textquotedblleft type
2\textquotedblright\ bubbles. Conditions for a bosonic stabilization of each
type of bubble due to a relativistic gas of bosonic particles have been determined.

\bigskip

\ \ \ It is found that type 1 bubbles can be metastable, finding an
equilibrium radius where the effects of the boson gas pressure balance those
of the wall tension, but this equilibrium exists for only a limited amount of
time. This is due to the fact that $\varphi$ particles with high enough
energy, $E\geq m$, can escape from the wall, and eventually wind up in the
exterior vacuum. The rate of leakage of boson particles is small initially,
but as the bubble shrinks, the gas temperature rises, and the rate of
$\varphi$ particle loss rapidly increases, leading to an unchecked collapse of
the type 1 bubbles.

\bigskip

\ \ Type 2 bubbles enclose a gas of $\phi$ bosons in the interior which exert
an outward pressure on the bubble wall. Again, the bubble can stabilize with
some radius $R$ when the gas has a temperature $T$. While all type 1 bubbles
equilibrate at the same temperature, regardless of size, a type 2 bubble
equilibrates at a temperature $T\propto R^{-1/4}$, so that a larger bubble at
equilibrium has a correspondingly lower temperature. Still, high energy $\phi$
particles can escape the bubble's interior by passing through the bubble wall,
but the reflection coeffficient $\mathcal{R}$ is energy dependent, with
$\mathcal{R}$ being large and the transmission coefficient $\mathcal{T}$
\ being small for low energies. Specifically, $\mathcal{\mathcal{R}}\approx1$
for particle energies $E\ll w^{-1}$, with $w$ being the thickness of the
bubble wall. For the bosonic stabilization mechanism to be long-lived, there
should be only a negligible fraction of $\phi$ particles with high enough
energy to escape, i.e., a very low leakage rate. The condition for this to
occur is found to be given by $T\ll w^{-1}$, so that at equilibrium, the
temperature is small in comparison to an energy $E_{c}\equiv w^{-1}$. For a
bubble near equilibrium, this implies $R\gg\Sigma w^{4}$, so that larger
equilibrium bubbles, or bubbles with very thin walls, may survive for longer
periods of time before an eventual decay.

\qquad


\begin{thebibliography}{99}
\bibitem {note}More specifically, we have $K=N_{0}m-\mathcal{E}_{0}$ for a
bubble near equilibrium, and $\mathcal{E}_{0}\approx\frac{4}{3}\mathcal{E}%
_{G}$, so that $K>0$ for $T\ll m$.
\end{thebibliography}
\end{document}